\documentclass[%
 reprint,
 amsmath,amssymb,
 aps,pre,longbiography
]{revtex4-2}

\usepackage{graphicx}
\usepackage{dcolumn}
\usepackage{bm}
\usepackage{paralist}
\usepackage{color}
\usepackage{physymb}
\usepackage{ulem}
\usepackage{url}

\begin{document}


\title{Anomalous Scaling of Aeolian Sand Transport Reveals Coupling to Bed Rheology}

\author{Katharina Tholen$^1$}
\author{Thomas P\"ahtz$^{2,3}$}
 \email{0012136@zju.edu.cn}
\author{Sandesh Kamath$^4$}
\author{Eric J. R. Parteli$^4$}
\author{Klaus Kroy$^1$}
 \email{klaus.kroy@uni-leipzig.de}
\affiliation{$^1$Institute for Theoretical Physics, Leipzig University, Postfach 100920, 04009 Leipzig, Germany}
\affiliation{$^2$Donghai Laboratory, 316021 Zhoushan, China}
\affiliation{$^3$Institute of Port, Coastal and Offshore Engineering, Ocean College, Zhejiang University, 316021 Zhoushan, China}
\affiliation{$^4$Faculty of Physics, University of Duisburg-Essen, Lotharstra{\ss}e 1-21, D-47057 Duisburg, Germany}

\date{\today}

\begin{abstract}
Predicting transport rates of windblown sand is a central problem in aeolian research, with implications for climate, environmental, and planetary sciences. Though studied since the 1930s, the underlying many-body dynamics is still incompletely understood, as underscored by the recent empirical discovery of an unexpected third-root scaling in the particle-fluid density ratio. Here, by means of grain-scale simulations and analytical modeling, we elucidate how a complex coupling between grain-bed collisions and granular creep within the sand bed yields a dilatancy-enhanced bed erodibility. Our minimal saltation model robustly predicts both the observed scaling and a new undersaturated steady transport state that we confirm by simulations for rarefied atmospheres. 
\end{abstract}

\maketitle
\begin{figure}[tb]
\includegraphics{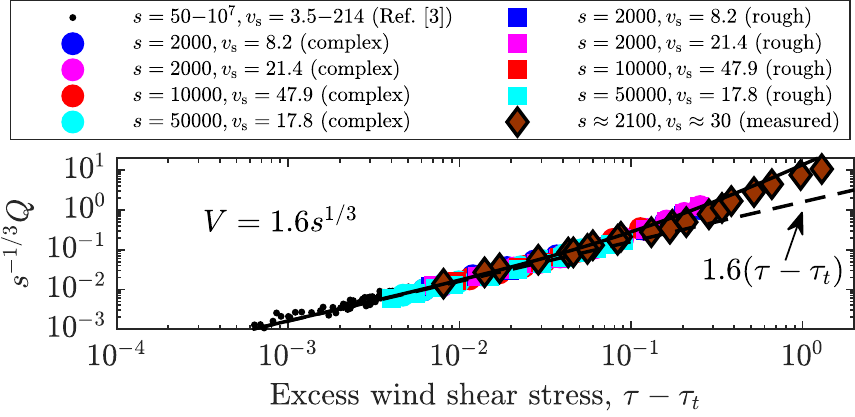}
\caption{Data from laboratory measurements \cite{Creysselsetal09,Hoetal11,Ralaiarisoaetal20} and previous \cite{PahtzDuran23} as well as our original sand-transport simulations, based on the discrete element method (DEM) \cite{Duranetal12,Kamathetal22} (see Supplementary Material \cite{SuppCooperativeSplash} for details), obey the transport-rate scaling in Eqs.~(\ref{Q}) and (\ref{V}) (solid line). The DEM simulations allow us to toggle between a complex boundary-layer wind velocity profile (dots and circles) and simplified ``fully rough’’ flow conditions (squares) based on Prandtl's turbulent closure \cite{Prandtl25}, cf.~Eq.~(\ref{ux}), and to study a wide range of particle-fluid density ratios $s$, terminal grain settling velocities $v_s$, and shear stresses $\tau$ in excess of the transport threshold $\tau_t$. The dashed line amounts to neglecting midair grain collisions.}
\label{LogLawScaling}
\end{figure}

Sand is a baffling material. It resembles a gas when shaken, a liquid when poured down a chute, and a solid when resting at a beach. When it is carried along by wind, all three manifestations are crucially involved side by side, making such aeolian transport a most revealing but also quite intricate sand-transport mode \cite{Bagnold41}. It is responsible for the spontaneous emergence of a multitude of granular surface waves in a variety of inorganic and organic sands, throughout the Solar System \cite{Hayes18}. Surprisingly, it also relates them to the drained halos that brighten up around your feet when you step on wet sand. To establish this  connection, we start from an empirically discovered scaling of the aeolian sand-transport rate $Q(s,v_s,\tau)$ as a function of the particle-fluid density ratio $s$ ($s\approx2100$ for quartz in air and $s\approx2.65$ in water), terminal grain settling velocity $v_s$, and wind shear stress $\tau$ \cite{PahtzDuran23}. In natural units, based on the grains’ median diameter and mass density, and the buoyancy-reduced gravitational acceleration $\tilde g\equiv(1-1/s)g$, it reads (Fig.~\ref{LogLawScaling})
\begin{subequations}
\begin{align}
 Q&=(\tau-\tau_t)[1+7.6(\tau-\tau_t)]V,\quad\text{with} \label{Q} \\
 V&=1.6s^{1/3}. \label{V}
\end{align}
\end{subequations}
This formulation splits the overall transport rate $Q$ into what is essentially the average velocity $V$ and density $\tau-\tau_t>0$ of mobilized grains \cite{PahtzDuran18b}. Intriguingly, the transport threshold $\tau_t(s,v_s)$ completely encapsulates the strength and functional form of fluid-particle interactions \cite{PahtzDuran23}. The usually subdominant term $7.6(\tau-\tau_t)$ is a semiempirical attempt to account for cooperative effects induced by intense winds, chiefly sand bed fluidization and midair grain collisions \cite{PasiniJenkins05,Carneiroetal13,Ralaiarisoaetal20,PahtzDuran20}. In the opposite limit, $\tau\approx\tau_t$, aeolian transport is idealized in terms of individual grains hopping along a static bed while dislodging additional grains, parametrized through a local ``splash function'' \cite{Beladjineetal07,Lammeletal17,Tanabeetal17,ComolaLehning17}, in the standard modeling approach \cite{Owen64,Kind76,Sauermannetal01,DoorschotLehning02,Sorensen04,Almeidaetal06,DuranHerrmann06,Almeidaetal08,Pahtzetal12,UngarHaff87,Creysselsetal09,JenkinsValance14,Berzietal16,JenkinsValance18,Andreottietal21,PahtzDuran20,Pahtzetal21, Huoetal21,Andreotti04,KokRenno09,Lammeletal12,LammelKroy17,Comolaetal22}. However, these conventional saltation models fail to recover Eq.~(\ref{V}), whose pure $s$ dependence and insensitivity to $v_s$ clashes with physical intuition and naive dimensional analysis \cite{PahtzDuran23}.

The primary objective of this Letter is to demonstrate a physical mechanism leading to such anomalous scaling. To this end, we first show that the mentioned failure of conventional saltation models is of general nature and hints at a coupling between the gaslike saltation layer and the rheology of the dense sand bed. The bed cannot be represented by a purely static granular packing, with a static-bed (local) splash function. Our discrete element method (DEM) simulations indeed reveal bed creep well below the yield point. While its direct contribution to $Q$ is negligible, bed creep and its concomitant nonlocal dilatancy cooperatively couple individual grain-bed collisions. Including this effect within a minimal analytical saltation model via a cooperative, dilatancy-enhanced  splash function indeed reproduces Eq.~(\ref{V}) and makes further testable predictions.

Consider a two-dimensional Cartesian coordinate system $(x,z)$, with wind direction $x$ and vertical direction $z$. For fluid-particle interactions via buoyancy and (for simplicity Stokes) drag, implying a terminal grain settling velocity $v_s=s/(18\nu)$ with the kinematic atmospheric viscosity $\nu$, the equations of motion for the $i$-th grain trajectory ($i=1,\dots,N$) read
\begin{subequations}
\begin{align}
 \dot v_z^i&=-1-v_z^i/v_s, \label{vz} \\
 \dot v_x^i&=(u_{x}-v_x^i)/v_s, \label{vx} \\
 \kappa^2(z+z_0)^2u_x^\prime|u_x^\prime|&=u_\ast^2[1-\tau_g(z)/\tau],&u_x(0)&=0.\label{ux}
\end{align}
\end{subequations}
The last equation is Prandtl's turbulent closure \cite{Prandtl25} for the wind velocity field $u_x(z)$ in the steady state, with the von K\'arm\'an constant $\kappa=0.4$, aerodynamic bed roughness $z_0=1/30$, wind shear velocity $u_\ast\equiv\sqrt{s\tau}$, and grain-borne shear stress profile $\tau_g(z)$. As the $xz$ component  of the granular stress tensor $(\sigma_{ij})$, the latter accounts for the streamwise momentum transfer between the wind and the grains along all grain trajectories: $\tau_g(z)=\sum_i\phi^i\Delta v_x^i(z)$. Here $\phi^i$ is the vertical flux of grains contributed by the $i$th trajectory and $\Delta v_x^i(z)$ the streamwise velocity gained between its ascending and descending visits of the elevation $z$. In the absence of grain motion ($\phi^1,\dots,\phi^N=0$), Prandtl's closure recovers the well-known (fully rough) law of the wall, $u_x=\kappa^{-1}u_\ast\ln(1+z/z_0)$. To close Eqs.~(\ref{vz})-(\ref{ux}), they are combined with a splash function, consisting of $2N$ boundary conditions linking the grain trajectories' impact velocities $\bm{v}_\downarrow^i$ to their lift-off velocities $\bm{v}_\uparrow^i$, and $N$ boundary conditions interconnecting the vertical flux contributions $\phi^i$. Importantly, for conventional, static-bed splash functions, all boundary conditions are fully determined by the impact velocities $\bm{v}_\downarrow^i$ \cite{Beladjineetal07,Lammeletal17,Tanabeetal17}. Hence, for given values of $v_s$ and $u_\ast$, the combined system of equations is closed and therefore has a fully determined solution $(\bm{v}^i,\phi^i/\tau)$. From this solution, all relevant global transport properties can be derived if also $s$ and thus $\tau=u_\ast^{2}/s$ are known. However, in blatant conflict with this analysis, $V$ in Eq.~(\ref{V}) is found to be independent of both $v_s$ and $u_\ast$, also in DEM simulations employing Prandtl's turbulent closure (Fig.~\ref{LogLawScaling}).

\begin{figure}[tb]
\includegraphics{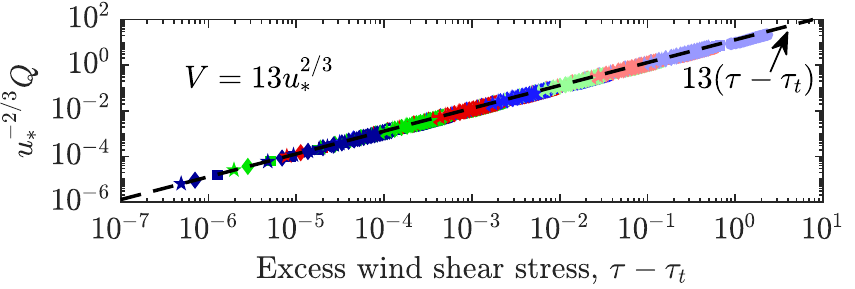}
\caption{Sand transport rate scaling predicted by our minimal two-species saltation model without midair collisions ($Q=(\tau-\tau_t)V$) and with a static-bed splash function \cite{Beladjineetal07} for terminal grain settling velocities $v_s=\{10^{3/2},10^2,10^{5/2},10^3\}$ (circles, squares, diamonds, stars) and particle-fluid density ratios $s=\{4^0,\dots,4^6\} \, v_s^2/10$ (colors). Small (large) $s$ tend to be on the right (left).}
\label{StaticBedScaling}
\end{figure}

Nonetheless, even the simplest nontrivial version of the above general model constitutes a minimal saltation model \cite{SuppCooperativeSplash} that can analytically reveal the origin of this discrepancy.  It combines the common \cite{Andreotti04,Lammeletal12} simplification of only considering two representative grain trajectories, namely high-energy saltons that rebound upon impact and their low-energy ejecta, the so-called reptons, with a closure mimicking the mass conservation found in the actual steady state \cite{SuppCooperativeSplash,Duranetal11}. With boundary conditions gleaned from an experimentally measured  splash function for a quiescent bed \cite{Beladjineetal07},  the calculated steady-state solutions $Q(\tau-\tau_t)$ (for saturated transport conditions) admit a data collapse consistent with $V=13u_\ast^{2/3}$ (Fig.~\ref{StaticBedScaling}), in line with previous observations based on (single- and multispecies) saltation models utilizing diverse static-bed splash functions \cite{Andreotti04,KokRenno09,Lammeletal12,LammelKroy17,Comolaetal22}. The scaling results from the height-dependent feedback of the grain trajectories on the wind. It seems, however, at odds with the widespread belief that the experimentally observed insensitivity of $V$ to the wind shear velocity $u_\ast$ is a consequence of the splash process \cite{Duranetal11,Koketal12}.

\begin{figure*}[tb]
\includegraphics{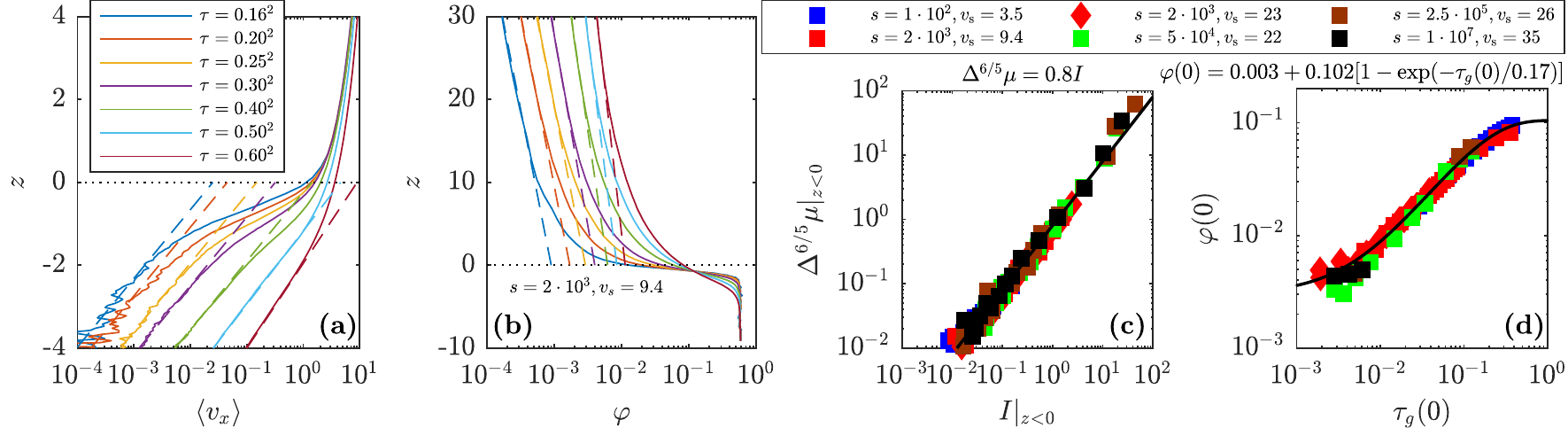}
\caption{Quasi-two-dimensional DEM-based sand-transport simulations (as in Ref.~\cite{PahtzDuran23}) of the creep and dilatancy regime of aeolian transport. (a) Granular creep visualized by the average height-resolved horizontal grain velocity $\langle v_x\rangle(z)$ in the sediment bed ($z<0$, below the elevation at which high-energy grain-bed collisions occur \cite{PahtzDuran18b,SuppCooperativeSplash}). Its increase with height $z$ and imposed wind shear stress $\tau$ (solid lines) reveals a characteristic skin depth $\lambda\approx0.72$ (dashed lines). (b) Because of progressive smoothing, the granular volume fraction profiles $\varphi(z)$ (solid lines) around $z=0$ deviate considerably from the limiting form for saltation on a quiescent bed---roughly a step from $\varphi\approx0.58$ to the exponential extrapolations of $\varphi(z>0)$ (dashed lines) \cite{Creysselsetal09,Hoetal11}. They exhibit a focal point $\varphi_f=\varphi(z\approx-\lambda)\approx0.1$. (c) The constitutive relation $\mu(\Delta,I)$ (solid line) for aeolian creep at subyield conditions ($\mu\lesssim0.3$) is similar to that of other sheared granular flows \cite{Gaumeetal20,KimKamrin20}. It interconnects the local friction coefficient $\mu=-\sigma_{xz}^c/\sigma_{zz}^c$, local normalized streamwise velocity fluctuations $\Delta\equiv(-T_{xx}/\sigma_{zz}^c)^{1/2}$ with $T_{xx}\equiv\langle v_x^2\rangle-\langle v_x\rangle^2$, and local inertial number $I\equiv\langle dv_x/dz\rangle/\sqrt{-\sigma_{zz}^c}$. Here, $\sigma_{xz}^c$ ($\sigma_{zz}^c$) is the shear (normal) component of the structural granular stress associated with grain-grain contacts in the bed \cite{SuppCooperativeSplash}. (d) The value $\varphi(z=0)$ is taken as a proxy for the number of grains available for splash ejection in Eq.~(\ref{Nephi}), and its $\tau_g(0)$ dependence motivates Eq.~(\ref{Ne}) with $\tau_Y\approx0.17$.}
\label{Creep}
\end{figure*}
\begin{figure*}[tb]
\includegraphics{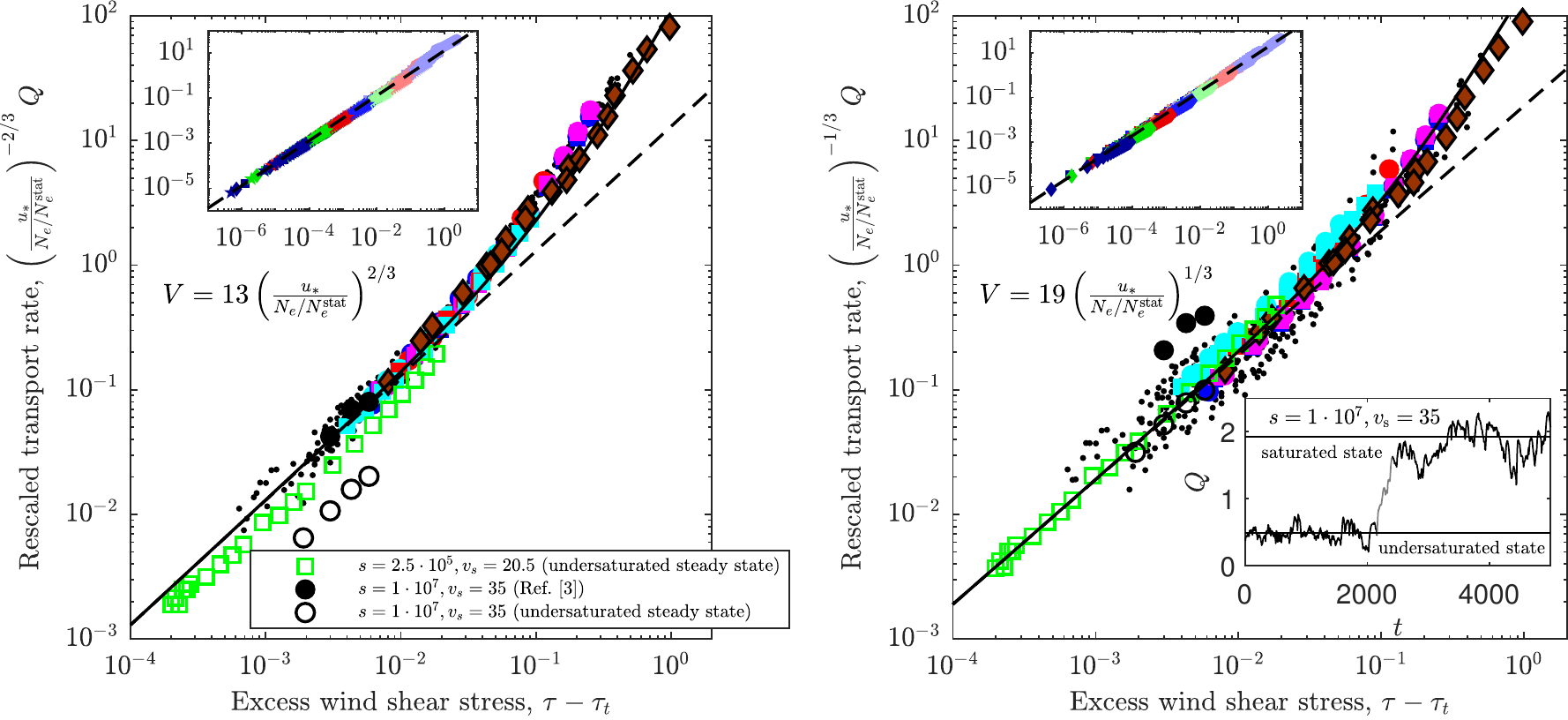}
\caption{Laboratory measurements, DEM-based sand-transport simulations (cf.~Fig.~\ref{LogLawScaling}), and predictions by our minimal saltation model with cooperative splash according to Eqs.~(\ref{Nephi}) and (\ref{Ne}) (inset) collapse on a master curve defined by (a) Eqs.~(\ref{Q}) and (\ref{V2}) (approximately $V\propto s^{1/3}$), corresponding to saturated transport conditions, or (b) an undersaturated steady state [Eqs.~(\ref{Q}) and (\ref{V3}), approximately $V\propto s^{1/6}$, upper inset]. Depending on the initial condition, this state can also be reached and sustained in DEM simulations, based on the code of Ref.~\cite{PahtzDuran23} (open black circles) or Ref.~\cite{Kamathetal22} (open green squares) for $s\gtrsim10^5$, regardless of the driving flow velocity profile (cf.~Fig.~\ref{LogLawScaling}). Lower inset: exemplary transition between the steady states, as occasionally spotted in the simulations. Solid (dashed) lines correspond to Eqs.~(\ref{Q}), (\ref{V}) with (without) the term representing midair collisions, which are neglected in our minimal saltation model. Filled symbols as in Figs.~\ref{LogLawScaling}, \ref{StaticBedScaling}.}
\label{MinimalModelScaling}
\end{figure*}

To resolve this apparent paradox, notice that, by dividing the right hand side of the relation $V=13u_\ast^{2/3}$ by $\tau^{1/3}=(u_\ast^2/s)^{1/3}$, one gets rid of the spurious $u_\ast$ dependence of $V$, and consistency with Eq.~(\ref{V}) is restored. While this procedure is inconsistent with the notion of a static-bed splash, we now show how it emerges by cooperative splash from a bed that is locally partially mobilized from earlier salton impacts. As revealed by Fig.~\ref{Creep}(a), the intermittent bed mobilization by impacting grains gives rise to a net granular creep upon averaging \cite{Pahtzetal20a}. The penetration of the emerging average grain velocity profile into the bed is characterized by a $\tau$-invariant skin depth $\lambda$ on the order of the grain diameter and associated with a considerable dilation of the bed, extending to a comparable depth [Fig.~\ref{Creep}(b)]. Additionally, our DEM simulations reveal an extended $\mu(I)$-rheological master relation \cite{Gaumeetal20,KimKamrin20} below the yield point [Fig.~\ref{Creep}(c)]. That it holds over a wide range of transport conditions establishes aeolian creep as a complex but well defined rheological phenomenology. Its robust constitutive law links the slow granular shearing motion driven by grain-bed collisions to the dissipation into (and the heating of) the bed. Its direct contribution to the overall transport rate $Q$ and momentum and energy dissipation is negligible---what matters is its indirect contribution via the dilatancy effect that enhances a subsequent splash and thereby boosts the highly dissipative repton layer \cite{Andreotti04}. 
 
To understand how this comes about, consider again Fig.~\ref{Creep}(b). For growing $\tau$, the step function of the granular volume fraction $\varphi(z)$ observed for a quiescent bed is increasingly smoothed, with an invariable focal point at $\varphi_f=\varphi(z\approx-\lambda)\approx0.1$. This is the dilatancy effect: a close-packed granular bed is jammed and cannot be sheared without dilating it to create free volume for the necessary grain rearrangements. It is the very mechanism that causes the aforementioned drainage and halos around the feet of  beach walkers \cite{Video}. As naturally expected, dilatancy affects the splash. In fact, recent DEM simulations have indicated an increase of the number $N_e$ of ejected bed-surface grains per salton with increasing impact frequency, while other splash properties such as the ejecta velocities remain nearly unaffected \cite{JiaWang22}. Since bed grains are effectively trapped (like in a Newton cradle), while hopping grains detach from their force chains, we assume that $N_e$ is directly proportional to the granular volume fraction $\varphi(0)$ at the rebound height $z=0$ (the ``mechanically pertinent bed surface,’’ $\lambda\approx0.72$ above the focal depth) \cite{PahtzDuran18b,SuppCooperativeSplash}:
\begin{equation}
 N_e/N_e^{\rm stat}=\varphi(0)/\varphi^{\rm stat}(0). \label{Nephi}
\end{equation}
This simple schematic model couples the gaslike layer of hopping grains above the bed surface to the dense-bed dynamics underneath and represents a crucial upgrade of the conventional static-bed splash parametrization, accounting for the dilatancy-mediated cooperativity. Remarkably, the observed splash geometry---in particular its characteristic surface radius $R=\mathcal{O}(10)$ \cite{Tanabeetal17,JiaWang22} and associated mobilized bed volume $N_e^{\rm stat}/\varphi^{\rm stat}\simeq 6 R^2\lambda=\mathcal{O}(600\lambda)$---is, together with $\varphi(0)\rightarrow\varphi^{\rm stat}(0)\approx3\times10^{-3}$ in the static-bed limit (Fig.~\ref{Creep}(d)), indeed consistent with the observation $N_e\rightarrow N_e^{\rm stat}=\mathcal{O}(1)$ \cite{Beladjineetal07,Lammeletal17,Tanabeetal17,ComolaLehning17}.  

Granular creep has been characterized as a sequence of stick-slip events, whereby slipping occurs when local fluctuations of the friction coefficient $\mu$ exceed the yield point \cite{Pahtzetal20a}. In our context of aeolian creep, characterized by its impact-induced local bed mobilizations with constant skin depth $\lambda=\mathcal{O}(1)$, $\mu$ reduces to the surface grain-borne shear stress $\tau_g(0)$ in our natural units \cite{Clarketal18}. Indeed, our DEM simulations show that Eq.~(\ref{Nephi}) is solely controlled by $\tau_g(0)$ via (Fig.~\ref{Creep}(d)):
\begin{equation}
 \varphi(0)=\varphi^{\rm stat}(0)+\varphi_f\left[1-\exp\left(-\tau_g(0)/\tau_Y\right)\right]. \label{Ne}
\end{equation}
The linear growth, $\varphi(0)=\varphi^{\rm stat}(0)+\varphi_f\tau_g(0)/\tau_Y$, for small $\tau_g(0)$ saturates near $\varphi_f$ (at $N_e=35N_e^{\rm stat}$) for large $\tau_g(0)$ [cf.~Fig.~\ref{Creep}(b)]. This suggests that the focal-point volume fraction $\varphi_f$ can be interpreted as the maximum $\varphi$ of fully mobile grains and therefore parametrizes a ``critical bed dilation,’’ below which bed force chains effectively disintegrate. The characteristic value $\tau_Y\approx0.17$, which determines both the linear increase and the saturation behavior in Eq.~(\ref{Ne}), can be linked to the yield friction $\mu_Y=\tau_Y/\varphi_b$ (for spheres, $\mu_Y\approx0.3$ \cite{Pahtzetal20a}) associated with an elementary yield event of a single bed grain at the static-bed volume fraction $\varphi_b\approx0.58$. In the same spirit, $\tau_Y/\varphi_f\approx1.7$ plays the role of a critical granular shear temperature required for grains to escape their traps and leapfrog over neighboring grains \cite{Creysselsetal09}.

As shown in Fig.~\ref{MinimalModelScaling}(a), data from our upgraded minimal saltation model, with cooperative splash according to Eqs.~(\ref{Nephi}), (\ref{Ne}), and $\tau_g(0)=\tau-\tau_t$ \cite{PahtzDuran18b}, collapse on
\begin{equation}
 V=13\left(\frac{u_\ast}{N_e/N_e^{\rm stat}}\right)^{2/3}, \label{V2}
\end{equation}
the master curve of the simulation and laboratory data. As expected, the transport threshold $\tau_t$ is not affected by this upgrade. The linear approximation of Eq.~(\ref{Ne}) with $1+(\tau-\tau_t)/\tau_e\approx2[(\tau-\tau_t)/\tau_e]^{1/2}$ (arithmetic mean $\approx$ geometric mean, where $\tau_e\equiv\tau_Y\varphi^{\rm stat}(0)/\varphi_f\approx5\times10^{-3}$) yields $V\approx1.4(1-\tau_t/\tau)^{-1/3}s^{1/3}$, deviating less than $13\%$ from Eq.~(\ref{V}) when $\tau/\tau_t\gtrsim2$. The anomalous scaling (compared to $V=13u_\ast^{2/3}$ for static-bed splash) has thus been traced back to the strongly skewed mass balance between reptons and saltons, originating from the creep-associated bed dilatancy. While their individual streamwise velocities exhibit the same increase with $u_\ast$ as in the static-bed case, the fraction of reptons increases by an order of magnitude with growing $\tau-\tau_t$, resulting in an almost $\tau$-invariant $V$.

Intriguingly, we moreover find that the steady-state condition in our minimal saltation model innately allows for an additional, undersaturated steady transport state [upper inset of Fig.~\ref{MinimalModelScaling}(b)], which scales as
\begin{equation}
 V=19\left(\frac{u_\ast}{N_e/N_e^{\rm stat}}\right)^{1/3}. \label{V3}
\end{equation}
Our DEM simulations indeed confirm its existence over a range of environmental conditions [Fig.~\ref{MinimalModelScaling}(b)]. For $s\lesssim10^5$, all simulations seem to approach the saturated steady state described by Eq.~(\ref{V2}), while some simulations for $s\gtrsim10^5$ can reach both steady states, Eq.~(\ref{V2}) or (\ref{V3}), for the explored initial conditions. Large random fluctuations can induce transitions between the steady states [lower inset of Fig.~\ref{MinimalModelScaling}(b)]. In view of the complexity of aeolian transport, the simultaneous quantitative agreement of both predicted steady states with grain-scale simulations provides strong support for our minimal two-species saltation model with cooperative splash.

In conclusion, we have shown that cooperative granular dynamics within the sand bed substantially affects aeolian sand-transport characteristics and can account for the anomalous scaling of the sand-transport rate $Q(s,v_s,\tau)$ [Eqs.~(\ref{Q}) and (\ref{V})]. The upshot is that grain-bed collisions cannot be portrayed as a sequence of isolated impacts on a purely static bed, but cooperate indirectly via the nonlocal and somewhat counterintuitive effect of creep-associated bed dilatancy. The main physical consequence is an increase of the relative population of (low-energy) reptating grains, which act as a momentum sink to the atmospheric boundary-layer flow. Our analytical two-species minimal saltation model, incorporating only a single representative salton and repton trajectory,  respectively,  identifies this cooperative, dilatancy-mediated negative feedback as the root cause behind the somewhat perplexing insensitivity of the average sand-transport velocity $V$ against substantial variations of the wind shear velocity $u_\ast$---thus challenging previous explanation attempts. Interestingly, it innately predicts an additional, undersaturated steady transport state, confirmed by our DEM simulations for conditions with extreme particle-fluid density ratio ($s\gtrsim10^5$), as typical for the thin atmospheres of Mars and Pluto. This calls for future studies of the competition between the two steady states in natural environments. It is also strongly indicative of the suitability of our analytical two-species saltation model for addressing the physical mechanism underlying other characteristic traits of aeolian transport.

\begin{acknowledgments}
This research was supported by a grant from the GIF, the German-Israeli Foundation for Scientific Research and Development (No.~155-301.10/2018). Furthermore, we acknowledge support from the National Natural Science Foundation of China (No.~12272344) and thank the German Research Foundation for funding through the Heisenberg Programme and the Grant No.~348617785. We particularly acknowledge the Regional Computing Center (RRZK) of the University of Cologne and the Centre for Information and Media Service (ZIM) of the University of Duisburg-Essen for computing time provided on the HPC systems CHEOPS and MagnitUDE, respectively.
\end{acknowledgments}

\renewcommand{\theequation}{S\arabic{equation}}
\renewcommand{\thefigure}{S\arabic{figure}}
\renewcommand{\thetable}{S\arabic{table}}

\section*{Supplementary Material}
\section{Natural units}
In the main text and this Supplemental Material, physical quantities are normalized using natural units, in terms of the particle density $\rho_p$, buoyancy-reduced gravity $\tilde g$, and median grain diameter $d$. For example, velocities are normalized by $\sqrt{\tilde gd}$.

\section{Numerical sand transport models}
The paper presents data from grain-scale discrete-element-method (DEM) simulations with the numerical models of Ref.~\cite{Duranetal12} (modified as described in Ref.~\cite{PahtzDuran17}) and Ref.~\cite{Kamathetal22}. Both models couple a continuum description of Reynolds-averaged aerodynamics with a discrete element method for the grain motion under gravity, buoyancy, and wind drag. The drag force is given by $\bm{F_d}=\frac{\pi}{8s}C_d|\bm{u_r}|\bm{u_r}$, where $\bm{u_r}$ is the fluid-grain-velocity difference and
\begin{equation}
 C_d=\left(\sqrt[m]{\frac{Re_c}{|\bm{u_r}|/\nu}}+\sqrt[m]{C_d^\infty}\right)^m
\end{equation}
the drag coefficient, with $[Re_c,C_d^\infty,m]=[24,0.5,2]$ in Ref.~\cite{Duranetal12} and $[Re_c,C_d^\infty,m]=[32,1,1.5]$ in Ref.~\cite{Kamathetal22}. Spherical grains ($10^4{-}10^5$) with mild polydispersity are confined in a quasi-two-dimensional \cite{Duranetal12} domain of length $\Delta_x\approx10^3$ and width $\Delta_y=1$ or three-dimensional domain of length $\Delta_x=200$ and width $\Delta_y=8$ \cite{Kamathetal22}, with periodic boundary conditions in the flow and lateral directions, and interact via normal repulsion and tangential friction. The grain layer at the bottom is glued to the ground. The upper boundary of the simulation domain is reflective but in practice never reached by transported grains. The Reynolds-averaged Navier-Stokes equations are combined with a semiempirical mixing-length closure,
\begin{equation}
 \frac{dl_m}{dz}=\kappa\left[1-\exp\left(-\sqrt{\frac{u_xl_m}{R_c\nu}}\right)\right], \label{lm}
\end{equation}
where $l_m(z)$ is the height-dependent mixing length, $\kappa=0.4$ is the von K\'arm\'an constant, $u_x(z)$ the flow velocity field, and $R_c=7$. It ensures a smooth hydrodynamic transition from high to low particle concentration near the bed surface and recovers the mean turbulent flow velocity profile in the absence of transport. The model of Ref.~\cite{Kamathetal22} allows to toggle between Eq.~(\ref{lm}) and a ``fully rough'' velocity profile, which neglects the bare viscous contribution to the wind shear stress against the turbulent contribution. Thereby, a simplified mixing length closure, $l_m=\kappa(z-z_s)$, becomes admissible, where $z_s$ is the uppermost height at which all particles move with a velocity smaller than $10\%$ of the wind shear velocity $u_\ast$. The integration of the velocity profile starts at $z=z_s+z_0$, where $z_0=1/30$ is the surface roughness. For $z\leq z_s+z_0$, $u_x(z)=0$.

\section{Calculation of physical quantities from the simulation data}
\subsection{Local averages}
Exploiting the spatial homogeneity of the simulation, we calculate the particle volume ($V_p$)-weighted average of a particle property $A_p$ over all particles within an infinitesimal vertical layer $(z,z+dz)$ and over all time steps after reaching the steady state as \cite{PahtzDuran18b}
\begin{equation}
	\langle A\rangle(z)=\sum_{z_p\in(z,z+dz)}V_pA_p/\sum_{z_p\in(z,z+dz)}V_p.
\end{equation}

\subsection{Particle volume fraction}
The particle volume fraction $\varphi$ is the total particle volume per simulation box volume $\Delta_x\times\Delta_y\times(z,z+dz)$ \cite{PahtzDuran18b}:
\begin{equation}
	\varphi(z)=\frac{1}{\Delta_x\Delta_ydz}\sum_{z_p\in(z,z+dz)}V_p.
\end{equation}
Note that the $\varphi$-values obtained from the quasi-two-dimensional simulations [Figs.~3(b) and 3(d)] are slightly different from those for three-dimensional systems (e.g., $\varphi$ is slightly smaller for a random closed packing). We expect that the relative deviations between quasi-two-dimensional and three-dimensional simulations are even smaller for $\varphi$-ratios such as in Eq.~(3).

\subsection{Sand transport rate}
The sand transport rate $Q$ is the total sand momentum per unit area of the bed and calculated as
\begin{equation}
	Q=\int_{-\infty}^\infty\varphi\langle v_x\rangle\mathrm{d}z.
\end{equation}

\subsection{Granular stresses}
The granular shear stress $\tau_g$ and pressure $p_g$ are calculated as \cite{PahtzDuran18b}
\begin{align}
 \tau_g&=\tau_g^k+\tau_g^c, \\
 p_g&=\varphi\langle v_z^2\rangle p_g^k+p_g^c,
\end{align}
where $\tau_g^k\equiv\sigma_{xz}^k=-\varphi\langle c_zc_x\rangle$ and $p_g^k\equiv-\sigma_{zz}^k=\varphi\langle c_z^2\rangle$, with $\bm{c}=\bm{v}-\langle\bm{v}\rangle$ the fluctuation velocity, are the shear and normal components, respectively, of the kinetic granular stress. Furthermore, denoting the contact force applied by grain $q$ on grain $p$ as $\mathbf{F}^{pq}$ (defining $\mathbf{F}^{pp}\equiv0$), the structural granular stress components $\tau_g^c\equiv\sigma_{xz}^c$ and $p_g^c\equiv-\sigma_{zz}^c$ are calculated as \cite{PahtzDuran18b}
\begin{align}
 \tau_g^c&=-\frac{1}{2\Delta_x\Delta_y}\sum_{z_p\in(z,z+dz)}\sum_qF_x^{pq}(z_p-z_q)l^{dz}_{pq}, \\
 p_g^c&=\frac{1}{2\Delta_x\Delta_y}\sum_{z_p\in(z,z+dz)}\sum_qF_z^{pq}(z_p-z_q)l^{dz}_{pq},
\end{align}
where $l^{dz}_{pq}$ the length fraction of the line connecting $z_p$ and $z_q$ that is contained in the interval $(z,z+dz)$.

\subsection{Bed surface}
The rebound height $z=0$, which serves as a pertinent mechanical definition of the notion of a ``bed surface'', is defined as the elevation at which $p_g^k\mathrm{d}\langle v_x\rangle/\mathrm{d}z$ is maximal:
\begin{equation}
	\left[p_g^k\frac{\mathrm{d}\langle v_x\rangle}{\mathrm{d}z}\right]_{z=0}=\max\left(p_g^k\frac{\mathrm{d}\langle v_x\rangle}{\mathrm{d}z}\right). \label{bedsurface}
\end{equation}
It is motivated by the balance equation \cite{Pahtzetal15a}
\begin{equation}
	-\frac{\mathrm{d}\varphi\langle c_z^2c_x\rangle}{\mathrm{d}z}=p_g^k\frac{\mathrm{d}\langle v_x\rangle}{\mathrm{d}z}-\varphi\langle a_xc_z\rangle-\varphi\langle a_zc_x\rangle,
\end{equation}
where $\bm{a}$ is the acceleration due to fluid-particle interactions, gravity, and contact forces, and $\bm{c}$ denotes the fluctuation velocity introduced above. Here $-\varphi\langle c_z^2c_x\rangle$ corresponds to the flux, $p_g^k\mathrm{d}\langle v_x\rangle/\mathrm{d}z$ to the production rate, and $\varphi\langle a_xc_z\rangle+\varphi\langle a_zc_x\rangle$ to the dissipation rate, respectively, of the kinetic fluctuation energy density $-\varphi\langle c_zc_x\rangle$. Since grain-bed rebounds produce strong correlations between the horizontal and vertical grain velocities, the maximum of $p_g^k\mathrm{d}\langle v_x\rangle/\mathrm{d}z$ corresponds to an effective rebound elevation \cite{PahtzDuran18b}. This definition provides the crucial link between saltation and bed mechanics and thereby establishes a one-to-one mapping between trajectory-based aeolian saltation models and simulations that explicitly resolve the granular bed. Note that, as a slight improvement beyond Ref.~\cite{PahtzDuran18b}, we insist that only the kinetic part of $p_g$ is used to calculate the rebound height in Eq.~(\ref{bedsurface}). This leads to a universal focal point at $\varphi_f\equiv\varphi(z\approx-0.72)\approx0.1$, intuitively consistent with the picture of grain reflection at $z\approx-\lambda$.

\section{Approximate analytical solution of trajectory equations} \label{TrajectoryEquations}
In this section, we present an approximate solution of Eqs.~(2a)-(2c), which will be used in the minimal model introduced in Section~\ref{MinimalModel}.

\subsection{Impact velocity as a function of lift-off velocity}
For a given wind velocity profile $u_x(z)$, the $i$-th grain's impact velocity $\bm{v}_\downarrow^i$ as a function of its lift-off velocity $\bm{v}_\uparrow^i$ can be calculated from Eqs.~(2a) and (2b) approximately as \cite{Pahtzetal21}
\begin{subequations}
\begin{align}
 v_{\downarrow z}^i&=-v_s-v_sW_0\left(-e^{-1-h^i/v_s^2}\right), \label{vz} \\
 v_{\downarrow x}^i&\simeq v_{\uparrow x}^i+\frac{v_{\uparrow z}^i-v_{\downarrow z}^i}{v_s+v_{\uparrow z}^i}[u_x(\overline{z}_\ast^i)-v_{\uparrow x}^i], \label{vx}
\end{align}
\end{subequations}
where $W_0$ denotes the principal branch of the Lambert-$W$ function, $\overline{z}_\ast^i$ is the $i$-th grain's characteristic transport height (a certain weighted average of $z^i$ \cite{Pahtzetal21}), and $h^i$ its hop height. They are calculated as
\begin{align}
 \overline{z}_\ast^i&=-(v_s+v_{\uparrow z}^i)v_{\downarrow z}^i-v_sv_{\uparrow z}^i, \\
 h^i&= v_sv_{\uparrow z}^i-v_s^2\ln(1+v_{\uparrow z}^i/v_s).
\end{align}
The approximation made in Eq.~(\ref{vx}), which assumes that $u_x[z^i(t)]$ changes much more slowly with time $t$ during the $i$-th grain's trajectory than $e^{t/v_s}$ (Eq.~(E2) in Ref.~\cite{Pahtzetal21}), is nearly exact for the log-like fully rough wind velocity profiles considered in Eq.~(2c).

\subsection{Wind velocity profile}
For $v_{\uparrow z}^i\lesssim v_s$, which holds for all our minimal model solutions, the difference between the $i$-th grain's downward and upward streamwise velocity $\Delta v_x^i(z)$ is approximately proportional to the vertical velocity difference $\Delta v_z^i(z)$ (Appendix). Approximating $\Delta v_z^i(z)$ to leading order in $v_{\uparrow z}^i/v_s$, the granular shear stress profile corresponding to the $i$-th trajectory is therefore given by
\begin{equation}
\begin{split}
 \tau_g^i(z)&\approx\tau_g^i(0)f_g(z/h^i),\quad\text{with} \\
 f_g(X)&\equiv\sqrt{1-X}\times\Theta\left(1-X\right),
\end{split} \label{taug}
\end{equation}
where $\tau_g^i(0)=\phi^i(v_{\downarrow x}^i-v_{\uparrow x}^i)$ and $\Theta$ denotes the Heaviside step function. When sorting the $N$ trajectories in ascending order of their hop heights ($h^1\leq h^2\dots\leq h^N$) and defining $h^0\equiv0$ and $\epsilon^i\equiv\tau_g^i(0)/\tau$, then the ratio between granular and wind shear stress $\tau_g(z)/\tau$ for elevations $h^{j-1}\leq z\leq h^j$ can be further approximated as
\begin{equation}
 \frac{\tau_g(z)}{\tau}\approx\sum_{i=j}^N\epsilon_i\sqrt{1-\frac{z}{h^i}}=\epsilon\left\langle\sqrt{1-\frac{z}{h^i}}\right\rangle_\epsilon\approx\epsilon\sqrt{1-\frac{z}{h}}, \label{taug2}
\end{equation}
where $\epsilon\equiv\sum_{i=j}^N\epsilon^i$, $h\equiv1/\langle(h^i)^{-1}\rangle_\epsilon$, with $\langle\cdot\rangle_\epsilon\equiv\frac{1}{\epsilon}\sum_{i=j}^N\cdot \,\epsilon^i$ the epsilon-weighted average over the trajectories $i=j,\dots,N$. In terms of the resulting wind velocity profile, we numerically confirmed that the combination of all of the above approximations leading to Eq.~(\ref{taug2}) typically cause deviations of less than $5\%$ from the exact solution even when $v_{\uparrow z}^i$ is on the order of $v_s$. These approximations have the advantage that they allow for an analytical solution of Prandtl's mixing length closure $\kappa^2(z+z_0)^2u_x^\prime|u_x^\prime|=u_\ast^2[1-\tau_g(z)/\tau]$. For elevations $h^{j-1}\leq z\leq h^j$, it reads
\begin{widetext}
\begin{equation}
\begin{split}
 u_x(z)&\approx\frac{u_\ast}{\kappa}\left\{\Re\left[f_u(h,z)\right]-\Im\left[f_u(h,z)\right]\right\}+u_x(h_{j-1}),\quad\text{with} \\
 f_u(h,z)&\equiv\left[4\sqrt{1-\epsilon f_g(z^\prime/h)}-2\sqrt{1+\epsilon\sqrt{2-f_g^2(z_0/h)}}\atanh\left(\frac{\sqrt{1-\epsilon f_g(z^\prime/h)}}{\sqrt{1+\epsilon\sqrt{2-f_g^2(z_0/h)}}}\right)\right. \\
 -2&{\left.\sqrt{1-\epsilon\sqrt{2-f_g^2(z_0/h)}}\atanh\left(\frac{\sqrt{1-\epsilon f_g(z^\prime/h)}}{\sqrt{1-\epsilon\sqrt{2-f_g^2(z_0/h)}}}\right)\right]^{z^\prime=z}_{z^\prime=h^{j-1}}+\ln\left(\frac{z+z_0}{h+z_0}\right)\times\Theta\left(\frac{z}{h}-1\right).} \label{ux}
\end{split}
\end{equation}
\end{widetext}
Using Eq.~(\ref{ux}), starting with the solution for $j=1$, the entire wind velocity profile can be analytically integrated, provided the grain trajectories and their vertical fluxes are known.

\section{Minimal model} \label{MinimalModel}
Our minimal saltation model considers two grain trajectories: high-energy \textit{saltons} (superscript $s$) that rebound indefinitely and eject low-energy \textit{reptons} (superscript $r$) upon impact with the bed. For simplicity, saltons are thus never captured by the bed, while reptons are captured after their first and only hop. The outcome of the grain-bed collisions is described by the static-bed splash function of Ref.~\cite{Beladjineetal07} (see below), optionally improved by Eq.~(4) to account for cooperative splash in the upgraded version of the model.

\subsection{Salton rebounds}
The rebound velocity $\bm{v}_{\uparrow}^s$ for a given impact velocity $\bm{v}_{\downarrow}^s$ of saltons is assumed to be equal to its experimentally measured average value, which can be empirically described by
\begin{subequations}
\begin{align}
 |\bm{v}_\uparrow^s|/|\bm{v}_\downarrow^s|&=A+Bv_{\downarrow z}^s/|\bm{v}_\downarrow^s|, \label{e} \\
 -v_{\uparrow z}^s/v_{\downarrow z}^s&=A/\sqrt{-v_{\downarrow z}^s/|\bm{v}_\downarrow^s|}-B, \label{ez}
\end{align}
\end{subequations}
where $A=0.87$ and $B=0.72$. Equation~(\ref{ez}) is a slight modification of the original empirical law given in Ref.~\cite{Beladjineetal07}. It reproduces the measurements \cite{Pahtzetal21} and, in contrast to the original law, respects the asymptotic scaling for small impact angles derived in Ref.~\cite{Lammeletal17}.

\subsection{Grain ejection}
The number $N_e^{\rm stat}=\phi^r/\phi^s$ of ejected reptons per impacting salton and their velocity $\bm{v}^r$ on a quiescent bed are assumed to be equal to their experimentally measured average values. Reference~\cite{Beladjineetal07} provided several empirical laws for these quantities, some in terms of the average impact and ejection velocities, others in terms of the average impact and ejection energies. Since static-bed splash is probably an energy balance problem \cite{Lammeletal17}, we choose the energy-based laws that are consistent with the assumption that the total ejection energy $N_e^{\rm stat}\bm{v}_\uparrow^{r2}$ is proportional to the rebound energy lost in the bed $\bm{v}_\downarrow^{s2}-\bm{v}_\uparrow^{s2}$:
\begin{subequations}
\begin{align}
 N_e^{\rm stat}&=13\left(1-\frac{\bm{v}_\uparrow^{s2}}{\bm{v}_\downarrow^{s2}}\right)\max\left(\frac{|\bm{v}_\downarrow^s|}{40}-1,0\right), \label{Ne0} \\
 \bm{v}_\uparrow^{r2}&=0.038(\bm{v}_\downarrow^{s2}-\bm{v}_\uparrow^{s2})/N_e^{\rm stat}, \label{ve} \\
 v_{\uparrow z}^{r2}/\bm{v}_\uparrow^{r2}&=30/38.
\end{align}
\end{subequations}
When cooperative splash is considered, the corrected number  $N_e$ of ejected reptons per salton impact is calculated from $N_e^{\rm stat}$ using Eqs.~(3) and (4). However, the calculation of $\bm{v}_\uparrow^{r2}$ does not change (i.e., $N_e^{\rm stat}$ in Eq.~(\ref{ve}) is not replaced by $N_e$).

\subsection{Steady state condition} \label{SteadyStateCondition}
In a multi-species trajectory-based model, the rare capture of saltons (the highest-energy grains) in the bed is, in the steady state, compensated by the promotion of a small fraction of reptons (the lowest-energy grains ejected by saltons) into saltons \cite{Andreotti04}, which is known as the replacement capacity condition \cite{Duranetal11}. Repton promotion is understood to be a dynamic process: the reptons gain hop by hop more and more energy, and this also decreases little by little their capture probability. For our minimalistic description, we neglect the capture of saltons and assume that reptons are always captured to keep the number of grain trajectories at $2$. However, this simplification causes the final system of equations (Section~\ref{SystemEquations}) to be underdetermined, resulting in a continuous phase space of steady-state solutions. We therefore need an additional selection criterion for a discrete subset of viable solutions,  analogous to the replacement capacity condition in more realistic multi-species systems. If reptons were allowed to perform multiple hops (contrary to our model assumption), it would mean that they would just fail to be promoted into saltons. To mimic this constraint, we require that the rebound energy of reptons in their (artificially suppressed) second hop would equal their ejection energy [cf. Eq.~(\ref{e})]:
\begin{equation}
 |\bm{v}_\uparrow^r|/|\bm{v}_\downarrow^r|=R(A+Bv_{\downarrow z}^r/|\bm{v}_\downarrow^r|), \label{R}
\end{equation}
with $R=1$.
\subsection{Solutions of system of minimal model equations} \label{SystemEquations}
\begin{figure}[tb]
\includegraphics{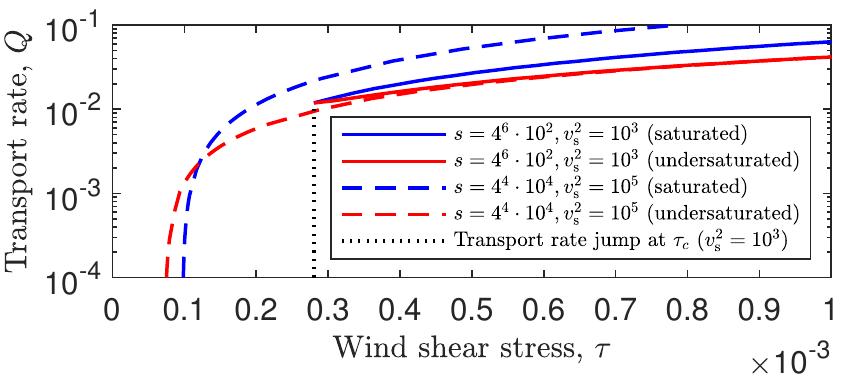}
\caption{Two alternative scenarios of the emergence of the saturated and undersatured steady states predicted by the minimal model (see Section~\ref{SystemEquations}).}
\label{Scenarios}
\end{figure}

\begin{figure}[tb]
\includegraphics{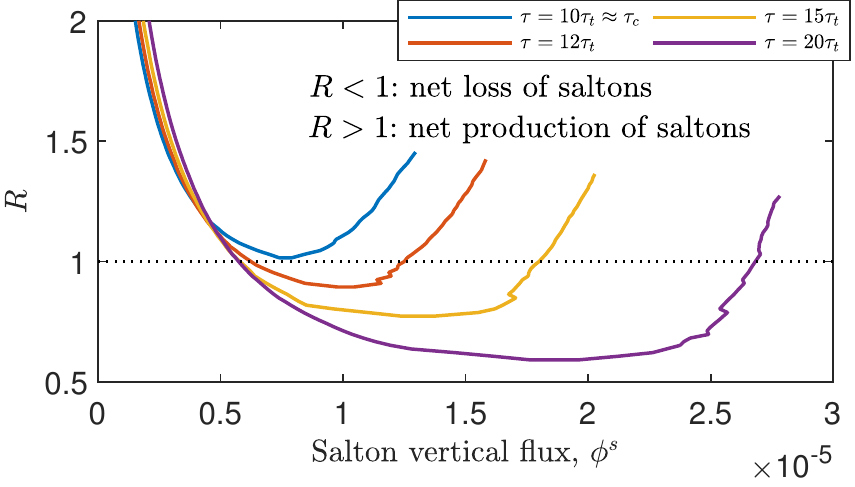}
\caption{Exemplary continuous phase lines of steady states allowed by the minimal model ($s=4^6\cdot10^2$ and $v_s^2=10^3$). For wind shear stress $\tau>\tau_c(s,v_s)$, the steady-state condition $R=1$ selects two steady-state solutions. The solution corresponding to the smaller value of the vertical salton flux $\phi^s$ is the saturated solution. It is stable against small perturbations along the phase line as $\mathrm{d}R/\mathrm{d}{\phi^s}<0$. The solution corresponding to the larger value of $\phi^s$ is the undersaturated, unstable ($\mathrm{d}R/\mathrm{d}{\phi^s}>0$) solution.}
\label{SteadyStates}
\end{figure}

The trajectory equations of Section~\ref{TrajectoryEquations} combined with the above boundary and steady-state conditions, optionally improved by Eq.~(4) to account for cooperative splash, constitute a closed system of equations, which we solve with an optimization algorithm. For given values of the control parameters $s$ and $v_s$, this system predicts two solutions when the wind shear stress $\tau$ is larger than a certain value $\tau_c(s,v_s)$: one corresponding to a saturated [inset of Fig.~4(a)] and one to an undersaturated steady state [upper inset of Fig.~4(b)]. When successively increasing $\tau$ from $0$ to $\tau_c(s,v_s)$, we observe two alternative scenarios (Fig.~\ref{Scenarios}). For the conditions $v_s^2 \in \{10^3,10^4\}$, both solutions simultaneously emerge (and coincide) at $\tau=\tau_c$. When crossing $\tau_c$, the transport rate $Q$ jumps from zero to the finite value $Q(\tau_c)$, implying that $\tau_c$ is larger than the transport threshold $\tau_t$ (e.g., Fig.~\ref{SteadyStates}), defined as the extrapolated value of $\tau$ at which $Q$ would vanish \cite{PahtzDuran18a}. For atmospheres with $v_s^2 \in \{10^5,10^6\}$, first the undersaturated state emerges at $\tau_t^{\rm uns}<\tau_c$ and then the saturated one at $\tau_t^{\rm sat}=\tau_c$,  for both states without jumps in $Q$. Hence, $\tau_t^{\rm sat}$ and $\tau_t^{\rm uns}$ are the distinct transport thresholds of the saturated and undersaturated solution, respectively. In the main text, the symbol `$\tau_t$' should be interpreted in the context of whether the undersaturated ($\tau_t=\tau_t^{\rm uns}$) or saturated state ($\tau_t=\tau_t^{\rm sat}$) is discussed.

\begin{figure}[tb]
\includegraphics{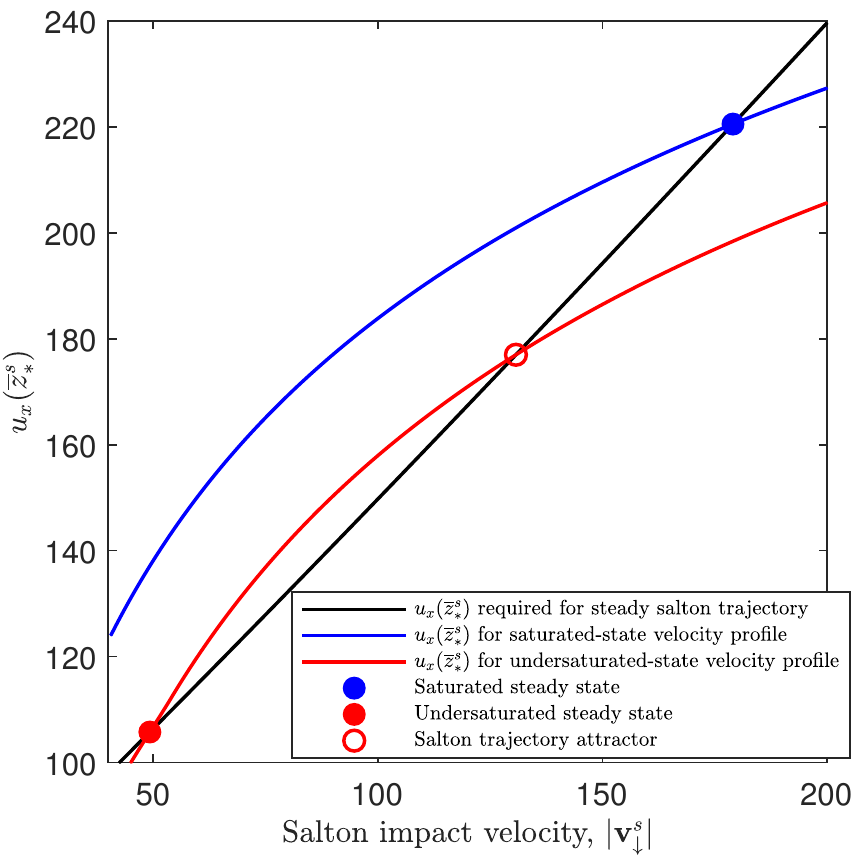}
\caption{Exemplary stability analysis for the salton trajectory ($s=4^4\cdot10^3$, $v_s^2=10^4$, and $\tau=8\tau_t$). Each steady state from the continuous phase line parametrized by $R$ (cf. Fig.~\ref{SteadyStates}) exhibits a certain impact velocity $|\bm{v}_\downarrow^s|$, characteristic transport height $\overline{z}_\ast^s$ of saltons, and associated mean wind velocity $u_x(\overline{z}_\ast^s)$ required to sustain it (black line). For the wind velocity profile of the saturated steady state (blue circle), $u_x(\overline{z}_\ast^s)$ (blue line) remains below (above) the black line for a positive (negative) perturbation of $|\bm{v}_\downarrow^s|$, keeping the salton trajectory stable. The opposite situation occurs for the undersaturated steady state (filled red circle) and its wind velocity profile (red line). Its salton trajectory is unstable and pulled towards a distant attractor (open red circle).}
\label{TrajectoryStability}
\end{figure}
\subsection{Stability of model solutions}
In order to analyze the stability of the two solutions, it is necessary to deactivate the steady-state condition and treat $R$ in Eq.~(\ref{R}) as a variable in the resulting continuous phase lines of steady states (Fig.~\ref{SteadyStates}). It describes how much salton capture is compensated by repton promotion, i.e., $R>1$ corresponds to net production and $R<1$  to a net loss of saltons. Hence, the function $R(\phi^s)$ does not only provide the two solutions corresponding to the steady-state condition ($R=1$) but also their stability against small perturbations along the phase line of steady states. Since $\mathrm{d}R/\mathrm{d}{\phi^s}<0\,(>0)$ for $\tau>\tau_c$, the (under)saturated solution at large (small) $|\bm{v}_{\downarrow}^s|$ and small (large) $\phi_s$ is (un)stable.  This is underscored by the following argument, venturing beyond the steady-state solutions of our minimal saltation model.   Consider a test salton that does not feed back onto the considered steady state of the minimal saltation model.  In the (under)saturated case this trajectory is (un)stable against small perturbations (Fig.~\ref{TrajectoryStability}). 

Unfortunately, these model implications seem to be at odds with our observation of apparently stable undersaturated steady states for certain conditions in the DEM simulations. To resolve the apparent contradiction, one would need to consider perturbations beyond the realm of steady states in a more consistent manner. This would require an elaborate analysis of the temporal response of the coupled fluid-particle system driven by fluid drag, salton capture, and repton promotion. Our simplistic two-species approximation is insufficient for such formal analysis, since repton promotion and salton capture are effectively encoded in the function $R(\phi^s)$ but not resolved in time. In the following, we give a flavor of how the dynamics of these processes can potentially stabilize the (nominally unstable) undersaturated state.

First, notice that, for perturbations of the undersaturated state along the phase line, increments of the salton velocity are only allowed in unison with a decrement of the salton vertical flux $\phi^s$. This is an unrealistic perturbation scenario, since capture of saltons is a result of their deceleration, not acceleration. On the one hand, if we only increment the velocity of the saltons, independent of $\phi^s$, their general tendency to further accelerate (Fig.~\ref{TrajectoryStability}) is counteracted by a stronger increase of the wind-grain momentum transfer as compared to perturbations along the phase line. This is due to the increase of  the repton vertical flux $\phi^r$, which dominates the overall momentum extraction from the wind. The effect is particularly pronounced in the undersaturated state, where $|\bm{v}_{\downarrow}^s|$ is typically close to the critical value $40$ in Eq.~(\ref{Ne0}), where $N_e$ vanishes. On the other hand, if we only decrement $\phi^s$, independent of the salton velocity, salton capture is counteracted by a weakening of the wind-grain momentum transfer, enhancing the promotion of reptons into saltons.  Again, this effect is particularly pronounced in the undersaturated state,  where repton promotion occurs very rapidly, since the ratio between the salton and repton energies is close to unity,  much smaller than in the saturated state. 
Taken together, these stabilizing effects of the wind-grain momentum transfer make it plausible that the undersaturated state may actually be stable against small perturbations for certain (if not all) conditions.  
In this case, the disparity of the two states would arguably increase with the density ratio $s$, in line with the observation that the undersaturated state occurs in the DEM simulations for the tested initial conditions only for large $s$.

\appendix*
\renewcommand{\theequation}{A\arabic{equation}}
\renewcommand{\thefigure}{A\arabic{figure}}
\renewcommand{\thetable}{A\arabic{table}}
\section{Grain velocity profile}
Equations~(\ref{vz}) and (\ref{vx}) are special cases of more general equations for the $i$-th grain's velocity $\bm{v}^i$ derived in Ref.~\cite{Pahtzetal21}, which can be written in the form
\begin{subequations}
\begin{align}
 v_z^i&=
  \begin{cases}
 -v_s-v_sW_{-1}\left(-e^{-1-(h^i-z)/v_s^2}\right)&\quad\text{if}\quad v_z^i\geq0\\
 -v_s-v_sW_0\left(-e^{-1-(h^i-z)/v_s^2}\right)&\quad\text{if}\quad v_z^i\leq0,
 \end{cases} \label{vzprofile} \\
 v_x^i&\simeq v_{\uparrow x}^i+\frac{v_{\uparrow z}^i-v_z^i}{v_s+v_{\uparrow z}^i}[u_x(z_{\uparrow\ast}^i)-v_{\uparrow x}^i], \label{vxprofileup}
\end{align}
\end{subequations}
where $W_n$ denotes the $n$-th branch of the Lambert-$W$ function and the characteristic transport height $z_{\uparrow\ast}^i$ is defined as
\begin{equation}
 z_{\uparrow\ast}^i\equiv2v_s^2+v_sv_{\uparrow z}^i+\frac{v_s(2v_s+v_z^i)(v_s+v_{\uparrow z}^i)}{v_{\uparrow z}^i-v_z^i}\ln\left(\frac{v_s+v_z^i}{v_s+v_{\uparrow z}^i}\right). \label{zup}
\end{equation}
Integrating the equations of motions backwards in time, starting with the $i$-th grain's impact velocity $\bm{v}_\downarrow^i$ as the initial condition, an alternative approximate expression for $v_x^i$ can be derived, analogous to Eq.~(\ref{vxprofileup}):
\begin{equation}
 v_x^i\simeq v_{\downarrow x}^i+\frac{v_{\downarrow z}^i-v_z^i}{v_s+v_{\downarrow z}^i}[u_x(z_{\downarrow\ast}^i)-v_{\downarrow x}^i], \label{vxprofiledown}
\end{equation}
with
\begin{equation}
 z_{\downarrow\ast}^i\equiv2v_s^2+v_sv_{\downarrow z}^i+\frac{v_s(2v_s+v_z^i)(v_s+v_{\downarrow z}^i)}{v_{\downarrow z}^i-v_z^i}\ln\left(\frac{v_s+v_z^i}{v_s+v_{\downarrow z}^i}\right). \label{zdown}
\end{equation}
The functions $z_{\uparrow\ast}^i(v_z^i)$ and $z_{\downarrow\ast}^i(v_z^i)$ intersect at a single point $v_{zc}^i$. In particular, they obey
\begin{equation}
 z_{\uparrow\ast}^i(v_{zc}^i)=z_{\downarrow\ast}^i(v_{zc}^i)=z_{\uparrow\ast}^i(v_{\downarrow z}^i)=z_{\downarrow\ast}^i(v_{\uparrow z}^i)=\overline{z}_\ast^i.
\end{equation}
Evaluating Eq.~(\ref{vxprofileup}) at $v_z^i=v_{\downarrow z}^i$ and Eq.~(\ref{vxprofiledown}) at $v_z^i=v_{\uparrow z}^i$ therefore yields
\begin{equation}
 \mu_b^i\equiv\frac{v_{\downarrow x}^i-v_{\uparrow x}^i}{v_{\uparrow z}^i-v_{\downarrow z}^i}\simeq\frac{u_x(\overline{z}_\ast^i)-v_{\uparrow x}^i}{v_s+v_{\uparrow z}^i}\simeq\frac{u_x(\overline{z}_\ast^i)-v_{\downarrow x}^i}{v_s+v_{\downarrow z}^i}. \label{mub}
\end{equation}
Furthermore, $z_{\uparrow\ast}^i(v_z^i)$ is a slowly varying function (by less than $36\%$) for $v_z^i\leq v_{zc}^i$ and typical values $v_{\uparrow z}^i<v_s$, and $z_{\downarrow\ast}^i(v_z^i)$ is a slowly varying function (by less than $13\%$) for $v_z^i\geq v_{zc}^i$ regardless of $v_{\uparrow z}^i$. Hence, using Eq.~(\ref{mub}), it follows that
\begin{equation}
 v_x^i\approx v_{\uparrow x}^i+\mu_b^i(v_{\uparrow z}^i-v_z^i)=v_{\downarrow x}^i+\mu_b^i(v_{\downarrow z}^i-v_z^i) \label{vxvzapprox}
\end{equation}
is a reasonable approximation for the entire range of $v_z^i$ and typical values $v_{\uparrow z}^i<v_s$. Equation~(\ref{vxvzapprox}) implies
\begin{equation}
 \frac{\Delta v_x^i(z)}{\Delta v_x^i(0)}\approx\frac{\Delta v_z^i(z)}{\Delta v_z^i(0)}.
\end{equation}
Its approximation to leading order in $v_{\uparrow z}^i/v_s$ (i.e., neglecting vertical drag), $\Delta v_z^i(z)=-2\sqrt{2(h^i-z)}$ [from Eq.~(\ref{vzprofile})], then implies Eq.~(\ref{taug}).

%

\end{document}